\documentclass[12pt]{article}

\usepackage{amsmath}
\usepackage{amsfonts}
\usepackage[body={17cm,23cm}]{geometry}
\begin{document}

\title{\bf  Axions and Their Relatives}
\author{ \bf{E. Mass{\'o}}\\
\small{ \it Grup de F{\'\i}sica Te{\`o}rica and Institut de F{\'\i}sica d'Altes Energies}\\
\small{  \it Universitat Aut{\`o}noma de Barcelona, 08193 Bellaterra, Spain}}

\date{Preprint UAB-FT-606}

\maketitle

\vspace*{.5cm}

\centerline{\large \bf Abstract}
A review of the status of axions and axion-like particles is given. Special attention is devoted to the recent results of the PVLAS collaboration, which are in conflict with the CAST data and with the astrophysical constraints. Solutions to the puzzle and the implications for new physics are discussed. The question of axion-like particles being dark matter is also addressed.

\vspace*{.5cm}

\section{The Axion}

In Quantum ChromoDynamics (QCD), we can use the gluon field strength $G^a$ to form the operator
\begin{equation}
 \theta \, {\alpha_s \over 8 \pi} \epsilon_{\mu\nu\rho\sigma} G^{a\mu\nu} G^{a\rho\sigma}
\label{masso_thetaterm}
\end{equation}
that has dimension four, and it is Lorentz and gauge invariant. We think this operator, named $\theta$-term, should be present in the QCD Lagrangian. In fact, it is thanks to the existence of the term (\ref{masso_thetaterm}) that we are able to solve the $U(1)_A$ problem, i.e., why $\eta'$ is heavy. The introduction of (\ref{masso_thetaterm}) in the QCD Lagrangian solves this problem but it introduces a new one, the strong CP problem. The $\theta$-term contributes to some physical observables, as for example the neutron electric dipole moment $d_n$. Experimentally there is a very tight upper limit
\begin{equation}
|d_n|\, < 0.6 \times 10^{-25}\, e\,{\rm cm}
\end{equation}
which implies the bound
\begin{equation}
| \theta + {\rm Arg\, Det}\, M| \, < \,  10^{-9}
\label{masso_boundtheta}
\end{equation}
In the lhs we see the two theoretical contributions to $d_n$.  Apart from the  $\theta$-term contribution, there is another related to
$M$, the quark mass matrix. The combination of the lhs is invariant under chiral rotations. In the rhs we see the number coming from the $d_n$ experiments.
We do not understand why the combination of the lhs is so tuned that leads to such a small number. This is the strong CP problem,  explained in detail in these Proceedings \cite{PecceiEM}.

An elegant solution to the strong CP problem was given by Peccei and Quinn
\cite{Peccei2EM}, based  on the hypothesis that
there is a new global $U(1)_{PQ}$ symmetry  that is spontaneously broken. How are we going to probe experimentally that this new symmetry exists? The task indeed seems hard. Fortunately, there is an unavoidable consequence of the Peccei-Quinn solution. The breaking of the $U(1)_{PQ}$ symmetry at an energy scale $f_a$ generates a Goldstone boson, that is called the axion. This fact
was realized in \cite{WeinbergEM}, where it was noticed that, more correctly, we have a pseudo Goldstone boson, in the sense that the global symmetry has a small explicit breaking due to quantum effects. This means a departure from a completely flat potential for the particle which in practical terms endows the axion with a calculable mass
\begin{equation}
 m_a   =
\frac{f_\pi m_\pi}{  f_a  } \frac{\sqrt{m_u m_d}}{m_u+m_d}=
0.6\ {\rm eV}\ \frac{10^7\, {\rm GeV}}{  f_a  }
\label{masso_massaxion}
\end{equation}
Here we have the quark masses $m_u$ and $m_d$ and the pion mass $m_\pi$, as well as the pion decay  constant $f_\pi= 93$ MeV. A consequence of (\ref{masso_massaxion}) is that once we fix the scale $f_a$ the mass of the axion $m_a$ is no longer a free parameter (and vice versa).

The interactions of the axion are model dependent. We are particularly interested in a coupling that is crucial for the current axion search experiments,
namely, the coupling of the axion field $a$ to two photons
\begin{equation}
 c_\gamma \, {\alpha \over 2 \pi\, f_a} \epsilon_{\mu\nu\rho\sigma} F^{\mu\nu}
F^{\rho\sigma} = - g_{a\gamma\gamma}\,   \vec E \vec B\, a
\label{masso_gammagamma}
\end{equation}
where $F$ is the electromagnetic field strength, and we have also written the interaction in terms of electric and magnetic fields. The model dependence is in the parameter $c_\gamma$ (or equivalently of course in
$g_{a\gamma\gamma}$). We can cook a model with fine tuned parameters and get a small $c_\gamma$, but in general $c_\gamma$ is of order one. For example in
the so called KSVZ axion type or ``hadronic axion'' \cite{KimEM}, where the axion
is not coupled to electrons at tree level, we have $c_\gamma=-0.97$. Another example is GUT embedded models like the so called DFSZ type \cite{DineEM}.
For the DFSZ type axion we have $c_\gamma=0.36$. In summary, we expect $c_\gamma= O(1)$, but we should keep in mind its (weak) model dependence. Other axions interactions are to gluons, which are model independent, and to matter. All the interactions are inversely proportional to $f_a$; we can see it explicitly in the $a\gamma\gamma$ case in
(\ref{masso_gammagamma}), i.e., $g_{a\gamma\gamma} \propto f_a^{-1}$. As we will see, the value of $f_a$  has to be quite large, and a consequence is that the axion is a very weakly interacting particle.

One finds constraints on the axion properties using
laboratory experiments as well as using astrophysical and cosmological
observations. There are  high energy laboratory experiments made in accelerators
that lead to $f_a>10^4$ GeV. However the strongest lower bounds on $f_a$ come from astrophysics. I briefly summarize them here; details can be found in these Proceedings \cite{RaffeltEM}.

Astrophysical limits are based on the idea that a ``too'' efficient energy drain due
to a possible stellar axion emission would change the time scale evolution of the star and would be inconsistent  with observation.
A very stringent limit using these ideas comes from horizontal branch stars in
globular clusters. The main production is from the Primakoff
process $\gamma \gamma^* \rightarrow a$ where $\gamma^*$ corresponds
to the electromagnetic field induced by protons and electrons in
the star plasma. The coupling is restricted to \cite{Raffelt2EM}
\begin{equation}
g_{a\gamma\gamma}   < 0.6 \times 10^{-10}\, {\rm GeV}^{-1} \hspace*{.3cm}
\Rightarrow \hspace*{.3cm}
  f_a   > 10^7\, {\rm GeV}
\label{masso_HB}
\end{equation}
However, the most restrictive astrophysical limits come from the analysis
of neutrinos from
SN 1987A. In the supernova core, the main axion production is
by bremsstrahlung in nucleon-nucleon processes.
The observed duration of the $\nu$ signal at the Earth detectors constrains
the coupling of the axion to nucleons and thus restricts the scale $f_a$
\cite{EllisEM}
\begin{equation}
f_a > 6 \times 10^8\, {\rm GeV}
\label{masso_limitSN}
\end{equation}

We should point out that when the scale $f_a<10^5$ GeV the SN constraints are no longer valid because the axions are so strongly coupled to nuclear matter that are trapped in the SN and do not stream freely from the core. However for such lower values of $f_a$ the other constraints, like (\ref{masso_HB}), are valid.  The conclusion is that (\ref{masso_limitSN}) is the lower bound on the Peccei-Quinn scale.

There are also cosmological arguments that put an upper bound to the scale $f_a$. During the Peccei-Quinn phase transition in the early universe, the axion field is a true Goldstone boson and picks up a vacuum expectation value. In the cooling of the universe the temperatures reach the QCD scale. Then a potential appears that drives the axion field to the CP conserving value. If the axion is too weakly coupled the
oscillation to the true minimum is so slow that the energy of the axion field could be greater than the critical density. This cannot happen of course, and the argument can be used to put an upper bound on $f_a$. This is the so called misalignment mechanism for axion production \cite{PreskillEM}. There are still other mechanisms, like axion production  from strings, although the studies that have been done about it do not coincide among themselves and there is no clear conclusion about the importance of the axion yield from strings.

As recent theoretical work on axions, we would like to mention \cite{Svrcek:2006yi},
where axions in string theory are re-examined, and \cite{Kaplan:2005wd}, where a model with an exotic axion cosmology is presented.

\section{The Axion Relatives}

Theories that go beyond the standard model of particle physics
have new symmetries, some of them global. Any time one of these
global symmetries is spontaneously broken we get
a Goldstone  or a pseudo Goldstone boson. An example is
family symmetry, which would be related to the
number and properties of families (we still do not have an answer to
Rabi's question:  Who ordered the muon?). The breaking of such a symmetry
would  give rise to familons. Another example is
lepton number symmetry, that would produce majorons. In general, in
theories beyond the $SU(3)\times SU(2)\times U(1)$ standard model there are quite often light scalar and pseudoscalar particles.
We will denote these new hypothetical light particles by $\phi$, and refer to them as axion-like particles (ALPs) \cite{Jaeckel:2006id}, both for the scalar and pseudoscalar case.

Quite generally the new particle  $\phi$ will couple to two photons. In the case that
$\phi$ is a pseudoscalar we have
\begin{equation}
{\cal
L}_{\phi\gamma\gamma}=\frac{1}{8M}
\epsilon_{\mu\nu\rho\sigma} F^{\mu\nu}
F^{\rho\sigma}\, \phi
 \label{masso_L}
\end{equation}
while for a scalar we would have
\begin{equation}
{\cal L}'_{\phi\gamma\gamma}=\frac{1}{4M}
 F_{\mu\nu}
F^{\mu\nu}\, \phi
\label{masso_LS}
\end{equation}
For both interaction lagrangians we have written the coupling as an inverse energy scale
$M$.

For a general ALP, there is no relation between mass and couplings, as there is for the axion as we see in (\ref{masso_massaxion}). In analyzing models with an ALP we will have two independent parameters: the mass $m$ of the light particle and the energy scale $M$ of new physics.

The reason for focusing on the coupling to two photons is that
most of the experiments that are searching for axions are based on such a coupling. This means that ALPs might induce a signal in such experiments.
Some of the bounds valid for axions are valid for an ALP. For example, the bound coming from the analysis of stellar energy losses in horizontal branch stars of globular clusters is based on the Primakoff production and thus applies. From
the upper limit on $g_{a\gamma\gamma}$ shown in (\ref{masso_HB}) one gets that
\begin{equation}
M >  10^{10} \ {\rm GeV} \label{masso_stellar}
\end{equation}
However, there are constraints that cannot be taken without modification. For example, the bound from the SN does not hold in the same way since in the axion model it uses the nucleon-axion coupling. One has to recalculate the SN limit when having only a $\phi\gamma\gamma$ coupling. While in realistic models the ALP usually couples to other particles, to assume that there is only a $\phi\gamma\gamma$ coupling, or that it dominates over other couplings, is the conservative option. The analysis of this scenario has been done in \cite{MassoEM},
and the conclusion is that the SN bound is looser than (\ref{masso_stellar}). Other constraints on ALPs  coupled to photons can be found in \cite{MassoEM, RabadanEM}; see also \cite{Rabadan2EM}.

Reciprocally, there are bounds that have no relevance for the invisible axion model but have their interest for ALPs.  Let us mention one example \cite{Grifols:1996id}. For very small mass of the ALP, it turns out the the emitted $\phi$ flux could be coherently reconverted to gamma rays in the galactic magnetic field.
Measurements on the SN1987A $\gamma-$ray flux by the Gamma-Ray Spectrometer on the Solar Maximum Mission satellite already imply a bound on the coupling $M > 3 \times 10^{11}$ GeV, valid for $m < 10^{-9}$ eV. It is exciting that the improved generation of satellite-borne detectors, like the project GLAST, could be able to detect a $\phi-\gamma$ signal from a nearby supernova, for allowed values of $M$.

Another aspect of interest is the possibility that ALPs could be the dark matter of the universe. We discuss it in Sect. \ref{DM_EM}.

Before finishing the Section, let us remark
that another reason to relax the relation (\ref{masso_massaxion}) is that
it could be no longer valid even in an axion model where there are
contributions to the axion mass from exotic sources \cite{Holman:1992us}.

\section{Searching for ALPs}

With the Peccei-Quinn scale $f_a$ pushed towards the high energy realm,  the axion becomes very light and extremely weakly interacting. The term "invisible axion" was coined, but Sikivie realized that it was not impossible to probe the existence of the
axion with feasible experiments \cite{SikivieEM}. There are several
ideas to look for axions,
all based on the coherent axion-photon conversion in a external strong magnetic field.

The ideas apply to any light ALP coupled to two photons. Indeed, we
can write the pseudoscalar case (\ref{masso_L}) as
\begin{equation}
{\cal
L}_{\phi\gamma\gamma}=\frac{1}{M}\, \vec E \vec B\,
\phi
\label{masso_L2}
\end{equation}
while for the scalar case (\ref{masso_LS}) can be written as
\begin{equation}
{\cal L}'_{\phi\gamma\gamma}=\frac{1}{2M}\, \left (
{\vec E}^2 - {\vec B}^2 \right)\, \phi
\label{masso_LS2}
\end{equation}
Both (\ref{masso_L2}) and (\ref{masso_LS2}) induce photon-ALP transitions in an external (classical) magnetic field.
In (\ref{masso_L2}), $B$ plays the role of the external magnetic field, and $E$ describes the photon field that couples to the field $\phi$.
In (\ref{masso_LS2}), one of the $B$ is again the external field, while it is now
the other $B$ that describes the photon field.

The photon-ALP mixing in a magnetic field
makes the interaction states $|\phi>$ and $|\gamma>$
different from the propagation states $|\phi'>$ and $|\gamma'>$,
\begin{eqnarray}
|\phi'> &=&      \cos\varphi\, |\phi>\, -\, \sin\varphi\, |\gamma> \\
|\gamma'> &=& \sin\varphi\, |\phi>\, +\, \cos\varphi\,  |\gamma>
\end{eqnarray}
The probability $P$ of the $\phi -\gamma$ transition has always
the suppression factor
factor $1/M^2$. However, the probability $P$
is enhanced when the $\phi -\gamma$ conversion in the magnetic
field is coherent.
A simple way to understand coherence is to describe
the photon and the axion as plane waves propagating
along a linear path of distance $L$. The conversion is coherent
provided there is overlap of the wave functions across a length
$L$, i.e.
\begin{equation}
|k_{\gamma'}-k_{\phi'}|\, L < 2\pi
\label{masso_coherent}
\end{equation}
In the coherent limit, the probability of the conversion is
\begin{equation}
  P (\gamma \rightarrow \phi) = \frac{1}{4}\,
  \frac{1}{M^2}\, \, B_T^2 \, L^2
\label{masso_P}
\end{equation}
Notice that only the transverse magnetic field $B_T$ is effective; this is easily understood from (\ref{masso_L2}) and (\ref{masso_LS2}).

We are interested in two methods to search for ALPs, that we describe now.

\subsection{Detection of solar ALPs}

There is a type of experiment, called helioscope \cite{SikivieEM}, that aims to detect ALPs from the Sun. It is
based on the fact that light particles coupled to photons would be produced in the
interior of the Sun, and subsequently leave it in the form of a
continuous flux. At the Earth, we can try to detect this solar flux
by looking at ALP back conversion to X-rays in a magnetic field.
The reason why the photon is in the X-ray range is because it carries the typical energies of the photons in the Sun interior, namely on the order of the keV.

There have been several helioscopes working in the last years, but without any doubt the helioscope of the CAST collaboration has reached an unprecedented level of accuracy. Until now they have not observed any signal, and this leads to the bound
\cite{ZioutasEM}
\begin{equation}
M > 0.9 \times 10^{10}  \ {\rm GeV}   \label{masso_cast}
\end{equation}
valid for  a mass of the light particle $m < 0.02$ eV. (See also the contribution to these Proceedings \cite{cast_EM})

\subsection{Production and detection of ALPs in the laboratory}

An helioscope is not thoroughly a laboratory experiment. Indeed, the assumed source of ALPs is the Sun. There is nothing wrong with this, after all we are able to detect solar neutrinos that have been produced in the solar interior. But it is desirable to have experiments where the ALP is produced and detected in terrestrial laboratories. An experiment with these characteristics was proposed in \cite{MaianiEM}. It consists in letting a polarized laser light to propagate in a magnetic field. The coupling of the ALP to photons makes possible the transition $\gamma \rightarrow \phi$, with the probability (\ref{masso_P}).

Notice that the absorption is selective. Take
the polarization  as the direction of the electric field of the laser beam.
In the case of a pseudoscalar $\phi$, from (\ref{masso_L2}) we see that it is the polarization parallel to $\vec B$ that is absorbed. In the scalar case, (\ref{masso_L2}) tells us that it is the polarization perpendicular to $\vec B$ that decreases. In both cases, the effect is a rotation of the plane of polarization. The selective absorption makes the vacuum dichroic in the presence of a magnetic field.

The PVLAS collaboration has been performing such experiment
\cite{ZavattiniEM} (see also \cite{Cantatore_EM}).
They do find a  signal
of rotation of the plane of the polarization of the laser. Their result can be interpreted
 in terms of an ALP. It is consistent with a scale
\begin{equation}
M \sim 4 \times 10^{5}\ {\rm GeV}  \label{masso_pvlas}
\end{equation}
and with a light particle mass
\begin{equation}
m \sim 10^{-3}\ {\rm eV}  \label{masso_m_pvlas}
\end{equation}

Before discussing in the next Section this result at the light of other bounds, we would like to comment on two further related issues. First, there is a second possible effect when light propagates in a magnetic field, as discussed in \cite{MaianiEM}. The double
virtual conversion $\gamma-\phi-\gamma$ produces a phase retardation of one of the polarizations. Which one is retarded depends on the parity of the particle $\phi$. This property manifests as vacuum birefringence, and it seems that there are positive results \cite{pc_EM}; we should take them as a preliminary result, but of course confirmation would be most exciting.

A final experiment we would like to mention is another (fully contained) laboratory method that could help in clarifying the puzzle we discuss in the next Section. It consists in the remarkable effect of light shining through a
wall: in a magnetic field, light oscillates into ALPs, these cross
a wall and afterwards they convert back into photons. This type of experiment was already made in \cite{CameronEM} with no signal observed that allowed to put some limits
- consistent with (\ref{masso_pvlas},\ref{masso_m_pvlas}). There are now several proposals for similar experiments  that are much more sensitive and that should  materialize in one or two years; they are explained in several contributions to these Proceedings \cite{everybodyEM}.

\section{Is it possible to evade the astrophysical constraints ?}

We ask this question because the PVLAS result (\ref{masso_pvlas},\ref{masso_m_pvlas}) strongly contradicts the astrophysical limits.
So, is there any ALP model where the astrophysical bounds are no longer valid?

I will try to convince the reader that the answer to this question is yes. But at the same time I think it is not an easy task. In the literature there are up to now only a handful of papers describing models that could be able to evade the astrophysical constraints on ALPs
\cite{Masso:2005ymEM,Masso:2006gcEM}. Hopefully, if the PVLAS results are confirmed the new physics will be restricted to a few selected models, either one of the list in
\cite{Masso:2005ymEM,Masso:2006gcEM}, or a new one. Apart from experiments, more theoretical work in this line is needed!

In this review I choose to summarize the paraphoton model of \cite{Masso:2006gcEM},
where it is assumed that the neutral $\phi$ particle couples to two
photons through a triangle diagram with an internal new fermion $f$.
We require two necessary conditions on the properties of $f$ and the triangle diagram.
The new particle $f$ should have a small electric charge on the
one hand, since it has to couple to photons, and on the other hand  this charge should decrease very sharply when
going from the momentum transfer involved in the PVLAS experiment,
$|k^2| \ll $ O(keV$^2$), to the typical momentum transfer in the solar
processes, $|k^2|\, \sim$ O(keV$^2$).

We can meet both conditions in the context of paraphoton
models \cite{Okun:1982xiEM,HoldomEM} if we introduce a very low energy scale. First,  as far as we know, these models are the
only ones where the effective electric charge of
some particles can be naturally very small. The idea is that
particles with a paracharge get an induced electric charge
proportional to some small mixing angle $\epsilon$ between photons
and paraphotons. To satisfy the second condition, i.e., to get a
variation of the effective electric charge with energy, we use a
model with two light but massive paraphotons having the same mixing
with the photon. If the fermion $f$ couples to the two paraphotons
with opposite paracharge, the resulting effective electric charge
for $f$ decreases with energy or temperature $T$,
\begin{equation}
q_f(T) \approx \frac{\mu^2}{T^2} q_f(0) \label{decrease}
\end{equation}
where $\mu$ is the mass scale of the paraphoton masses. In
(\ref{decrease}) we have assumed $T\gg\mu$. With the low energy scale $\mu \simeq 10^{-3}$ eV
and $\epsilon$ such that $q(0)e \simeq 10^{-8}e$, the model is able
to accommodate the strength of the PVLAS signal and yet have a very
suppressed emission in the Sun. Notice that in this paraphoton model the CAST
limit (\ref{masso_cast}), which is based on a standard solar $\phi$-flux,
does not hold.

Concerning the nature of $\phi$, there are three possibilities

\begin{description}

\item [i)] $\phi$ is a fundamental particle, coupled to $f$,

\item [ii)] $\phi$ is a $f\bar f$ composite particle, with $f$ and $\bar f$ confined by new forces,

\item [iii)] $\phi$ is not really a particle, it is positronium-like state. In the same way positronium is a $e^- e^+$ bound state,  $\phi$ is a  $f \bar f$ bound state.

\end{description}

The PVLAS results have stimulated some new lines of thought. For
example, an explanation of the data not in terms of light particles coupled to photons is given in \cite{Mendonca:2006pgEM}. Alternative axion models that could make compatible all observations are worked out in \cite{Antoniadis:2006wpEM}. Finally, I would like to mention the work in \cite{Gies:2006caEM}, where it is shown that the  results from PVLAS and from other related experiments can be used to bound the properties of epsilon-charged particles.

\section{ALPs as dark matter}\label{DM_EM}

We have many independent indications of the existence of dark matter (DM): from the local measurements of the galactic rotation curves to the joint fit of the high redshift SN data and of the anisotropies of the Cosmic Microwave Background (CMB). Also, we have several motivated DM candidates. Only experiment will tell us in the future which one is the selected, although we should keep in mind that quite possibly there might be several different components contributing to the DM of the universe.

Among the popular candidates for DM we have the axions. With this motivation there are on going searches that probe the possible contribution of axions to the galactic DM
\cite{van_bibberEM}. In this section we would like to discuss some aspects about the possibility that ALPs might be DM.

A first result is the following. Assume the existence of a light ALP that is coupled {\em only} to two photons, either (\ref{masso_L}) or
(\ref{masso_LS}), but {\em not} to other particles. In this case, the final result is that the ALP cannot be DM. The reason is that this particle is born thermally. The ALP species was in thermal equilibrium in the early universe thanks to the photon coupling. There is a moment where the expansion rate of the universe is greater than the interaction rate so that the $\phi\gamma\gamma$ coupling is no longer effective to maintain equilibrium. The relic number of ALPs today is less than the number of relic photons today in the CMB.
It follows that only when they have masses on the order of the keV, ALPs could in principle contribute substantially to the DM. However, with such masses, ALPs are unstable  due to the decay $\phi \rightarrow \gamma \gamma$. Thus, as we said,
the result is negative: an ALP with only a photon coupling cannot be DM.

In fact, leaving apart the DM issue, there are even more restrictions.
Consider the region of the parameter space $(m,M)$ that would correspond to a relic density not far from the critical density of the universe if we ignore decays. In the realistic case of decays, the region is excluded by the constraints coming from CMB distortion and He photodissociation  \cite{MassoEM}.

 This negative conclusion may be altered in realistic models of ALPs with couplings  to matter, and also where we should take into account  the possibility of other mechanisms of generation of ALPs in the early universe, as happens with the axions.

An example of a model with ALPs that are not only coupled to photons and where the ALPs may be DM is presented in \cite{rotaEM}.
The model describes the effects of a small explicit breaking of a global symmetry, as suggested by gravitational arguments. It has one scalar field transforming under a global U(1) symmetry, and coupled to matter and to gauge bosons. The spontaneous breaking of the explicitly broken symmetry gives rise to a massive pseudo Goldstone boson, i.e, to an ALP. In such a model one analyzes thermal and non-thermal production of ALPs in the early universe, and performs a systematic study of astrophysical and cosmological constraints on the ALP properties. The conclusion is that for very suppressed explicit breaking the pseudo Goldstone boson is a cold dark matter candidate  \cite{rotaEM}. Such a suppression is not unexpected according to some analyses of gravitational symmetry breaking  \cite{Kallosh:1995hiEM}.

\section{Conclusion}

The experimental search for light particles coupled to photons, that we call axion-like particles (ALPs), is reaching an unprecedented level of sensitiveness. There are already some results (from the PVLAS collaboration) that may be interpreted in terms of ALPs. If confirmed, we have to search for models that make compatible their apparent inconsistency  with the sound astrophysical bounds and with other results, for example from the CAST collaboration.
We have examined these issues. Also, we have reviewed the possibility that ALPs may contribute to the dark matter present in our universe.

\section*{Acknowledgments}

 I would like to thank   B. Beltr{\'a}n, M. Kuster, T. Papaevengelou and  K.  Zioutas for the successful axion meeting they organized at CERN.
Also, thanks go to my colleagues J. Redondo, A. Ringwald, J. Jaeckel and F. Takahashi for many discussions that have helped me in the writing of this  contribution.
Finally, I acknowledge
support by the CICYT Research Project FPA2005-05904 and the
\textit{Departament d'Universitats, Recerca i Societat de la
Informaci{\'o}} (DURSI), Project 2005SGR00916.


\begin{thebibliography}{99}

\bibitem{PecceiEM}
See R.~D.~Peccei, these Proceedings

\bibitem{Peccei2EM}
R.~D.~Peccei and H.~R.~Quinn,
Phys.\ Rev.\ Lett.\  {\bf 38},  1440 (1977)
\\
R.~D.~Peccei and H.~R.~Quinn,
Phys.\ Rev.\ D {\bf 16},  1791 (1977)


\bibitem{WeinbergEM}
S.~Weinberg,
Phys.\ Rev.\ Lett.\  {\bf 40},  223 (1978)
\\
F.~Wilczek,
Phys.\ Rev.\ Lett.\  {\bf 40},  279 (1978)


\bibitem{KimEM}
J.~E.~Kim,
Phys.\ Rev.\ Lett.\  {\bf 43},  103 (1979)
\\
M.~A.~Shifman, A.~I.~Vainshtein and V.~I.~Zakharov,
Nucl.\ Phys.\ B {\bf 166},  493 (1980)

\bibitem{DineEM}
M.~Dine, W.~Fischler and M.~Srednicki,
Phys.\ Lett.\ B {\bf 104}, 199  (1981)
\\
A.~R.~Zhitnitsky,
Sov.\ J.\ Nucl.\ Phys.\  {\bf 31},  260 (1980)
[Yad.\ Fiz.\  {\bf 31},  497 (1980)]

\bibitem{RaffeltEM}
See G.~G.~Raffelt, these Proceedings

\bibitem{Raffelt2EM}
G.~G.~Raffelt,
``Stars As Laboratories For Fundamental Physics:
The Astrophysics Of Neutrinos, Axions, And Other
Weakly Interacting Particles,''
{\it  Chicago, USA: Univ. Pr. (1996) 664 p}.

\bibitem{EllisEM}
J.~R.~Ellis and K.~A.~Olive,
Phys.\ Lett.\ B {\bf 193}, 525  (1987)\\
G.~Raffelt and D.~Seckel,
Phys.\ Rev.\ Lett.\  {\bf 60},  1793 (1988)\\
M.~S.~Turner,
Phys.\ Rev.\ Lett.\  {\bf 60}, 1797  (1988)

\bibitem{PreskillEM}
J.~Preskill, M.~B.~Wise and F.~Wilczek,
Phys.\ Lett.\ B {\bf 120},  127 (1983)
\\
L.~F.~Abbott and P.~Sikivie,
{\it ibid.}, 133
\\
M.~Dine and W.~Fischler,
{\it ibid.}, 137
\\
M.~S.~Turner,
Phys.\ Rev.\ D {\bf 33}, 889  (1986)

\bibitem{Svrcek:2006yi}
  P.~Svrcek and E.~Witten,
  JHEP {\bf 0606}, 051 (2006)
  [arXiv:hep-th/0605206]\\
P.~Svrcek,
  arXiv:hep-th/0607086

\bibitem{Kaplan:2005wd}
  D.~B.~Kaplan and K.~M.~Zurek,
  Phys.\ Rev.\ Lett.\  {\bf 96}, 041301 (2006)
  [arXiv:hep-ph/0507236]

\bibitem{Jaeckel:2006id}
  J.~Jaeckel, E.~Masso, J.~Redondo, A.~Ringwald and F.~Takahashi,
  arXiv:hep-ph/0605313

\bibitem{MassoEM}
  E.~Masso and R.~Toldra,
  Phys.\ Rev.\ D {\bf 52}, 1755 (1995)
  [arXiv:hep-ph/9503293];
  Phys.\ Rev.\ D {\bf 55}, 7967 (1997)
  [arXiv:hep-ph/9702275]

\bibitem{RabadanEM}
  M.~Kleban and R.~Rabadan,
  arXiv:hep-ph/0510183

\bibitem{Rabadan2EM}
  R.~Rabadan,
  these Proceedings

\bibitem{Grifols:1996id}
  J.~A.~Grifols, E.~Masso and R.~Toldra,
  Phys.\ Rev.\ Lett.\  {\bf 77}, 2372  (1996)
  [arXiv:astro-ph/9606028];\\
J.~W.~Brockway, E.~D.~Carlson and G.~G.~Raffelt,
  Phys.\ Lett.\ B {\bf 383}, 439 (1996)
  [arXiv:astro-ph/9605197]

\bibitem{Holman:1992us}
R.~Holman, S.~D.~H.~Hsu, T.~W.~Kephart, E.~W.~Kolb, R.~Watkins and L.~M.~Widrow,
Phys.\ Lett.\ B {\bf 282}, 132 (1992)
[arXiv:hep-ph/9203206];\\
M.~Kamionkowski and J.~March-Russell,
Phys.\ Lett.\ B {\bf 282}, 137 (1992)
[arXiv:hep-th/9202003];\\
S.~M.~Barr and D.~Seckel,
Phys.\ Rev.\ D {\bf 46}, 539 (1992)

\bibitem{SikivieEM}
  P.~Sikivie,
  Phys.\ Rev.\ Lett.\  {\bf 51}, 1415 (1983)
  [Erratum-ibid.\  {\bf 52}, 695 (1984)]


\bibitem{ZioutasEM}
  K.~Zioutas {\it et al.}  [CAST Collaboration],
  Phys.\ Rev.\ Lett.\  {\bf 94}, 121301 (2005)
  [arXiv:hep-ex/0411033]

\bibitem{cast_EM}
See CAST collaboration, these Proceedings

\bibitem{MaianiEM}
  L.~Maiani, R.~Petronzio and E.~Zavattini,
  Phys.\ Lett.\ B {\bf 175}, 359 (1986)


\bibitem{ZavattiniEM}
  E.~Zavattini {\it et al.}  [PVLAS Collaboration],
  Phys.\ Rev.\ Lett.\  {\bf 96}, 110406 (2006)
  [arXiv:hep-ex/0507107]

\bibitem{Cantatore_EM}
See G. Cantatore, these Proceedings

\bibitem{pc_EM}
G. Cantatore and U. Gastaldi, private communication

\bibitem{CameronEM}
  R.~Cameron {\it et al.},
  Phys.\ Rev.\ D {\bf 47}, 3707  (1993)

\bibitem{everybodyEM}
See A. Ringwald, P. Pugnat, etc., these Proceedings

\bibitem{Masso:2005ymEM}
  E.~Masso and J.~Redondo,
  JCAP {\bf 0509}, 015 (2005)
  [arXiv:hep-ph/0504202];\\
  P.~Jain and S.~Mandal,
  arXiv:astro-ph/0512155;\\
  J.~Jaeckel et al., in   \cite{Jaeckel:2006id}

\bibitem{Masso:2006gcEM}
  E.~Masso and J.~Redondo,
  arXiv:hep-ph/0606163

\bibitem{Okun:1982xiEM}
L.~B.~Okun,
Sov.\ Phys.\ JETP {\bf 56}, 502 (1982) [Zh.\ Eksp.\ Teor.\ Fiz.\
{\bf 83}, 892 (1982)]

\bibitem{HoldomEM}
B.~Holdom,
Phys.\ Lett.\ B {\bf 166}, 196 (1986);
Phys.\ Lett.\ B {\bf 178}, 65 (1986)

\bibitem{Mendonca:2006pgEM}
  J.~T.~Mendonca, J.~Dias de Deus and P.~Castelo Ferreira,
   ``Higher harmonics in non-linear vacuum from QED effects without low mass
  arXiv:hep-ph/0606099

\bibitem{Antoniadis:2006wpEM}
  I.~Antoniadis, A.~Boyarsky and O.~Ruchayskiy,
  arXiv:hep-ph/0606306

\bibitem{Gies:2006caEM}
  H.~Gies, J.~Jaeckel and A.~Ringwald,
  arXiv:hep-ph/0607118

\bibitem{van_bibberEM}
See K. van Bibber, these Proceedings

\bibitem{rotaEM}
  E.~Masso, F.~Rota and G.~Zsembinszki,
  Phys.\ Rev.\ D {\bf 70}, 115009  (2004)
  [arXiv:hep-ph/0404289]

\bibitem{Kallosh:1995hiEM}
  R.~Kallosh, A.~D.~Linde, D.~A.~Linde and L.~Susskind,
  Phys.\ Rev.\ D {\bf 52}, 912  (1995)
  [arXiv:hep-th/9502069]


\end{thebibliography}
\end{document}